\documentclass[A4,12pt]{article}
\usepackage{amsmath,cite,epsfig}

\def\ti              {\tilde}
\def\wt              {\widetilde}

\def\b               {\beta}

\def\l               {\lambda}

\def\x               {\chi}

\def\st              {\ti t}
\def\sb              {\ti b}
\def\stau            {\ti \tau}
\def\snu             {\ti \nu}

\def\ch              {\ti \x^\pm}
\def\nt              {\ti \x^0}
\def\sg              {\ti g}

\newcommand{\mst}[1]   {m_{\ti t_{#1}}}
\newcommand{\msb}[1]   {m_{\ti b_{#1}}}
\newcommand{\mstau}[1] {m_{\ti\tau_{#1}}}

\newcommand{\mnt}[1]   {m_{\ti \x^0_{#1}}}

\newcommand {\iosqrttwo} {\frac{i}{\sqrt{2}}}
\newcommand {\oosqrttwo} {\frac{1}{\sqrt{2}}}

\def \sT {\widetilde T}
\def \sB {\widetilde B}
\def \sTc {\widetilde T^c}
\def \sBc {\widetilde B^c}

\def \sf {\tilde f}
\def \st {\tilde t}
\def \sb {\tilde b}
\def \stau {\tilde\tau}
\def \Rf {R^{\tilde f}}
\def \Rt {R^{\tilde t}}
\def \Rb {R^{\tilde b}}
\def \Rfc {R^{\tilde f\,*}}
\def \Rtc {R^{\tilde t\,*}}
\def \Rbc {R^{\tilde b\,*}}

\def \muff {\mu_{e\!f\!f}}

\def \gev {~{\rm GeV}}

\newcommand{\eq}[1]  {\mbox{(\ref{eq:#1})}}
\newcommand{\fig}[1] {Fig.~\ref{fig:#1}}
\newcommand{\Fig}[1] {Figure~\ref{fig:#1}}
\newcommand{\tab}[1] {Table~\ref{tab:#1}}

\newcommand{\gsim}{\;\raisebox{-0.9ex}
           {$\textstyle\stackrel{\textstyle >}{\sim}$}\;}

\newcommand{\lsim}{\;\raisebox{-0.9ex}{$\textstyle\stackrel{\textstyle<}
           {\sim}$}\;}

\begin{document}

\begin{flushright}
   \vspace*{-18mm}
   CERN-PH-TH/2005-117 \\
   IFIC/05-30 \\
   ZU-TH 11/05
\end{flushright}
\vspace*{2mm}

\begin{center}

{\LARGE  Sfermion decays into singlets and singlinos\\[3mm]
         in the NMSSM } \\[10mm]

{\large  S. Kraml$^{\,1}$, W. Porod$^{\,2,3}$ }\\[4mm]

{\it 1) CERN, Dep. of Physics, Theory Division, Geneva, Switzerland\\[1mm]
     2) Instituto de F\'isica Corpuscular, Universitat de Val\`encia, Spain\\[1mm]
     3) Institut f\"ur Theoretische Physik, Univ. Z\"urich, Switzerland }

\end{center}

\begin{abstract}
We investigate how the addition of the singlet Higgs field 
in the NMSSM changes the sfermion branching ratios as compared to the MSSM. 
We concentrate in particular on the third generation, 
discussing decays of the heavier stop, sbottom or stau 
into the lighter mass eigenstate plus a scalar or pseudoscalar singlet Higgs. 
We also analyse stop, sbottom and stau decays into singlinos. 
It turns out that the branching ratios of these decays can be large,  
markedly influencing the sfermion phenomenology in the NMSSM. 
Moreover, we consider decays of first and second generation sfermions 
into singlinos.
\end{abstract}

\section{Introduction}

The Next-to-Minimal Supersymmetric Standard Model (NMSSM) provides an 
elegant solution to the $\mu$ problem of the MSSM by the addition of 
a gauge singlet superfield $\hat S$ \cite{Fayet:1974pd}. 
The superpotential of the Higgs sector then has the form 
$\lambda\hat S(\hat H_d\cdot\hat H_u)+\frac{1}{3}\kappa\hat S^3$. 
When $\hat S$ acquires 
a vacuum expectation value, this creates an effective $\mu$ term, 
$\mu\equiv\lambda\langle S\rangle$, which is automatically of the 
right size, {\it i.e.}\ of the order of the electroweak  scale. 
In this way, in the NMSSM the electroweak scale originates entirely 
from the SUSY breaking scale.

The addition of the singlet field leads to a larger particle spectrum 
than in the MSSM: in addition to the MSSM fields, the NMSSM contains two 
extra neutral (singlet) Higgs fields --one scalar and one pseudoscalar-- 
as well as an extra neutralino, the singlino. Owing to these extra states, 
the phenomenology of the NMSSM can be significantly different from the MSSM; 
not at least because the usual LEP limits do not apply to the singlet 
and singlino states. 

The NMSSM has recently become very popular. 
Most of the Feynman rules are given in \cite{Franke:1995tc}. 
The NMSSM Higgs phenomenology has been investigated extensively in 
\cite{Ellwanger:1993hn,Elliott:1993uc,Ellwanger:1995ru,King:1995vk,Ellwanger:1999ji,Ellwanger:2001iw,%
Miller:2003ay,Ellwanger:2003jt,Ellwanger:2004gz,Miller:2004uh,%
Ellwanger:2005uu,Ellwanger:2005fh} (a model variant without a $\hat S^3$ 
term is discussed in \cite{Panagiotakopoulos:2000wp}). 
Detailed studies of the neutralino sector are available in 
\cite{Franke:1994hj,Franke:1995tf,Ellwanger:1997jj,Hesselbach:2000qw%
,Hesselbach:2001ri,Franke:2001nx,Choi:2004zx,Moortgat-Pick:2005vs}. 
The relic density of (singlino) dark matter has been studied  
in \cite{Belanger:2005kh}.
The sfermion sector, on the other hand, has so far received very little 
attention, although here, too, one may observe differences 
as compared to the MSSM. 
In this letter we therefore investigate the decays of squarks and sleptons 
in the framework of the NMSSM and contrast them to the MSSM case.
We concentrate on the third generation (stops, sbottoms and staus) where 
we expect the largest effects, but also consider decays of selectrons,
smuons and 1st/2nd generation squarks. 

In the MSSM, squarks and sleptons can decay via $\sf_i^{}\to f\nt_k$, 
$\sf_i^{}\to f'\ch_j$ with $i,j=1,2$ (or $i=L,R$ in case of no mixing) 
and $k=1,...,4$. 
Squarks can also decay into gluinos, $\ti q_i^{}\to q\sg$, if the gluino is 
light enough. In addition, sfermions of the third generation 
($\sf=\st,\sb,\stau$) can have the bosonic decay modes 
$\sf_i^{}\to\sf_j'+W^\pm\!,\,H^\pm$
and $\sf_2^{}\to\sf_1^{}+Z^0\!,\,h^0\!,\,H^0\!,\,A^0$ \cite{Bartl:1998xk}.  
In the NMSSM, we have additional decay modes into singlinos and singlet Higgs 
states: $\sf_i^{}\to f\nt_n$ with $n=1,...,5$ and 
$\sf_2^{}\to\sf_1^{} + A_1^0,\,A_2^0,\,H_1^0,\,H_2^0,\,H_3^0$. 
Pure singlets and singlinos couple in general 
very weakly to the rest of the spectrum. There are hence two potentially 
interesting cases: a) large mixing of singlet and doublet Higgs states 
and/or large mixing of singlinos with gauginos-higgsinos and 
b) (very) light singlet/singlino states.  
In case b) $A_1^0$ and $H_1^0$ are almost pure 
singlets with masses well below the LEP bound of $m_h\ge 114$~GeV, 
and the singlino is the LSP. 
We investigate these cases in this letter and show that they can markedly 
influence the sfermion phenomenology. 

The paper is organized as follows. 
In Section~2 we explain our notation, the potential 
and the relevant Feynman rules. 
In Section~3 we perform a numerical analysis, and in 
Section~4 we present our conclusions.

\section{Notation and couplings}

\subsection{Potential}

We follow the notation of NMHDECAY \cite{Ellwanger:2004xm}.
The superpotential is then given as 
\footnote{Note the different signs of the $h_b$ and $h_\tau$ terms as 
compared to Eq.~(A.1) of Ref.~\cite{Ellwanger:2004xm}.}$^,$\footnote{The 
superpotential Eq.~(\ref{eq:superpot}) posesses a discrete 
$Z_3$ symmetry which is spontaneously broken at the electroweak 
phase transition. This results in cosmologically dangerous domain 
walls \cite{Abel:1995wk}. 
We implicitly assume a solution \cite{McDonald:1997vy} to this 
domain wall problem which does not impact collider phenomenology.}
\begin{eqnarray}
  {\cal W} = h_t\, \hat Q \cdot \hat H_u \hat T^c_R 
    + h_b\, \hat H_d \cdot \hat Q  \hat B^c_R
    + h_\tau\, \hat H_d \cdot \hat L\,\hat\tau^c_R
    - \lambda \hat S\, \hat H_d \cdot \hat H_u
    + \frac{1}{3} \kappa \hat S^3
\label{eq:superpot}
\end{eqnarray}
with the $SU(2)$ doublet superfields 
\begin{eqnarray}
   \hat Q = {{\hat T}_L \choose {\hat B}_L} \,,\quad 
   \hat L = {{\hat \nu}_L \choose {\hat \tau}_L} \,,\quad 
   \hat H_u = {{\hat H}_u^+ \choose {\hat H}_u^0} \,,\quad 
   \hat H_d = {{\hat H}_d^0 \choose {\hat H}_d^-} 
\end{eqnarray}
and the product of two $SU(2)$ doublets  
\begin{eqnarray}
  \hat X_1 \cdot \hat X_2 = \hat X^1_1 \hat X^2_2 - \hat X^2_1 \hat X^1_2 \,.
\end{eqnarray}
From Eq.~(\ref{eq:superpot}) we derive the $F$-terms  
\begin{align}
   F_{T_L}    &=\,   h_t \sTc_R H^0_u - h_b \sBc_R H^-_d \,, & 
   F_{T^c}    &=\,   h_t (\sT_L H^0_u - \sB_L H^+_u) \,,\\
   F_{B_L}    &=\, - h_t \sTc_R H^+_u + h_b \sBc_R H^0_d \,, & 
   F_{B^c}    &=\, - h_b (\sT_L H^-_d - \sB_L H^0_d) \,,\\
   F_{\tau_L}\, &=\,   h_\tau \wt\tau^c_R H^0_d \,, & 
   F_{\tau^c}\, &=\, - h_\tau (\wt\nu_L H^-_d - \wt\tau_L H^0_d) \,,\\
   F_{H^0_u}  &=\,   h_t \sT_L \sTc_R - \lambda S H^0_d \,, & 
   F_{H^+_u}  &=\, - h_t \sB_L \sTc_R + \lambda S H^-_d \,,
\label{eq:FHu} \\
   F_{H^0_d}  &=\,   h_b \sB_L \sBc_R + h_\tau\wt\tau_L^{}\wt\tau^c_R - \lambda S H^0_u \,, & 
   F_{H^-_d}  &=\, - h_b \sT_L \sBc_R - h_\tau\wt\nu_L^{}\wt\tau^c_R + \lambda S H^+_u \,,
\label{eq:FHd}
\end{align}
yielding the Yukawa part of the scalar potential 
${\cal V}_F = \sum_i F_i F^*_i$. Note that the $F$-terms in Eqs.~(\ref{eq:FHu})
and (\ref{eq:FHd}) imply direct interactions between the $S$ field and the 
sfermions:
\begin{eqnarray}
{\cal V}_F &\supseteq& - h_t \lambda^* \left( 
        \sB_L \sTc_R S^* {H^-_d}^* +  \sT_L \sTc_R S^* {H^0_d}^* \right) 
 - h_b \lambda^* \left( 
        \sB_L \sBc_R S^* {H^0_u}^* + \sT_L \sBc_R S^* {H^+_u}^*
             \right) \nonumber \\
 && - h_\tau \lambda^* \left( 
        \stau_L \stau_R^c S^* {H^0_u}^* + \snu_L \stau_R^c S^* {H^+_u}^*
             \right)  + {\rm h.c.} \,\,.
\label{eq:VF}
\end{eqnarray}
We also need the soft SUSY-breaking potential for the derivation of the 
couplings, 
c.f.\ Eq.~(A.4) of \cite{Ellwanger:2004xm},  
\begin{eqnarray}
  {\cal V}_{soft} = 
    h_t A_t \wt Q \cdot  H_u \wt T^c_R 
  + h_b A_b  H_d \cdot \wt Q \wt B^c_R
  + h_\tau A_\tau  H_d \cdot \wt L \wt\tau^c_R
  - \lambda A_\lambda  S \,  H_d \cdot  H_u
  + \frac{1}{3} \kappa A_\kappa S^3 \,.
\end{eqnarray}

\subsection{Sfermion--Higgs interaction}

In the following, 
we denote the neutral scalar and pseudoscalar Higgs bosons by 
$H^0_i$ ($i=1,2,3$) and $A^0_l$ ($l=1,2$), respectively.
The interaction of $H^0_i$ and $A^0_l$ with a pair of sfermions 
$\tilde f_j^{}\tilde f_k^*$ ($j,k=1,2$) can be written as:
\begin{eqnarray}
  {{\cal L}}_{{\ti f\ti f\phi}} &=&   
         g^S_{ijk}\, H^0_i \tilde f_j^{}\tilde f^*_k 
       + g^P_{ljk}\, A^0_l \tilde f_j^{}\tilde f^*_k\,.
\end{eqnarray}
Apart from $D$-term contributions, the Higgs--sfermion couplings 
are proportional to the Yukawa couplings $h_f$. We therefore write 
$g^{S,P}_{ijk}$ explicitly for the third generation. 
For stops, we have 
\begin{eqnarray}
(g^S_{ijk})^{\st} &=& 
      \left(\oosqrttwo\, h_t\,  
         (\muff^* \Rtc_{j1}\Rt_{k2} + \muff \Rtc_{j2}\Rt_{k1} )
                          - v_d D_{jk} \right) S_{i2} \nonumber \\
   & & - \left( \oosqrttwo\, h_t\, 
         (A_t \Rtc_{j1}\Rt_{k2} + A^*_t \Rtc_{j2}\Rt_{k1} )  
         - v_u D_{jk} 
         + \sqrt{2} v_u h^2_t\delta_{jk}  \right) S_{i1} \nonumber \\
   & & +\, \oosqrttwo \, v_d h_t  \,
         (\lambda^* \Rtc_{j1} \Rt_{k2}
          + \lambda \Rtc_{j2} \Rt_{k1} )\, S_{i3} \,,  \\[3mm]
(g^P_{ljk})^{\st} &=& -\, \iosqrttwo\, h_t\,
     ( \muff^* P_{l2} +  A_t P_{l1} +  v_d \lambda^*  P_{l3} )
     \Rtc_{j1} \Rt_{k2} 
 \nonumber \\ &&
   +\, \iosqrttwo\, h_t\, 
    ( \muff P_{l2} +  A^*_t P_{l1} +  v_d \lambda  P_{l3} )
    \Rtc_{j2} \Rt_{k1}  \,,  \label{eq:gPljkst}
\end{eqnarray}
where $\muff$ is the effective $\mu$ term:
\begin{eqnarray}
  \muff \equiv \lambda s 
\end{eqnarray}
with $s=\langle S\rangle$ the vev of the singlet $S$. 
(In the presence of an additional generic $\mu$ term $\mu \hat H_d\hat H_u$,  
$\muff\to \muff = \lambda s + \mu$.) 
For sbottoms, we get 
\begin{eqnarray}
(g^S_{ijk})^{\sb} &=& -
 \left( \oosqrttwo\,  h_b\, ( A_b \Rbc_{j1} \Rb_{k2}
                         + A^*_b \Rbc_{j2} \Rb_{k1} )
              + v_d D_{jk}
          +  \sqrt{2} v_d h^2_b  \delta_{jk}  \right) S_{i2} \nonumber \\
   & &  + \left( 
     \oosqrttwo\, h_b\,  ( \muff^* \Rbc_{j1} \Rb_{k2}
                     + \muff \Rbc_{j2} \Rb_{k1} )
                    + v_u D_{jk} \right) S_{i1} \nonumber \\ 
  & & +\, \oosqrttwo \,
   v_u h_b\, (\lambda^* {\Rb_{j1}}^*\Rb_{k2} + \lambda \Rbc_{j2}\Rb_{k1} ) 
    \,S_{i3}  \,, \\[3mm]
(g^P_{ljk})^{\sb} &=& -\, \iosqrttwo\,h_b \,
     ( A_b P_{l2} + \muff^* P_{l1} + v_u \lambda^* P_{l3} )  
     \Rbc_{j1} \Rb_{k2}
 \nonumber \\ &&
 +\, \iosqrttwo\, h_b\, 
     ( A^*_b P_{l2} + \muff P_{l1} +  v_u \lambda  P_{l3} )
     \Rbc_{j2} \Rb_{k1}  \,,  \label{eq:gPljksb}
\end{eqnarray}
and analogously for staus with the obvious replacements $h_b\to h_\tau$, 
$A_b\to A_\tau$ and $R^{\sb}\to R^{\stau}$. 

In Eqs.~(16)--(19), $S_{ij}$ and $P_{ij}$ are the Higgs mixing matrices 
as in \cite{Ellwanger:2004xm}, 
and $\Rf$ are the sfermion mixing matrices diagonalizing the sfermion 
mass matrices in the notation of \cite{Bartl:2003pd,Gajdosik:2004ed}:  
\begin{eqnarray}
   {\sf_1\choose\sf_2} = \Rf\, {\sf_L\choose\sf_R}, \quad 
   {\rm diag}(m_{\sf_1}^2,\, m_{\sf_2}^2)
   = \Rf \left(\begin{array}{cc} 
        m_{\sf_L}^2 & a_f^* m_f \\ a_f m_f & m_{\sf_R}^2
     \end{array}\right) (\Rf)^\dagger
\end{eqnarray}
where $a_t = A_t - \muff^*\cot\beta$ and 
$a_{b,\tau} = A_{b,\tau} - \muff^*\tan\beta$.  
Furthermore, $v_d=\langle H_d^0\rangle$, $v_u=\langle H_u^0\rangle$ and 
$\tan\beta=v_u/v_d$. The $D$-terms are given by
\begin{eqnarray}
D_{jk}^{\tilde f} &=& \frac{1}{\sqrt 2} \left( 
                 (g^2 T_{3L} - Y_L {g'}^2) \Rfc_{j1} \Rf_{k1} 
                          - Y_R  {g'}^2 \Rfc_{j2} \Rf_{k2}  \right) \,
\end{eqnarray}
where $g$ and $g'$ are the $SU(2)$ and $U(1)$ gauge couplings:  
$g^2=4\sqrt{2} G_F M_W^2$ with $G_F$ the Fermi constant and 
${g'}^2=g^2\tan^2\theta_W=4\sqrt{2} G_F (M_Z^2-M_W^2)$ using the 
on-shell relation $\sin^2\theta_W=(1-M_W^2/M_Z^2)$ for the Weinberg 
angle $\theta_W$; 
$T_{3L}$ and $Y_{L(R)}$, are the 3rd component of the isospin 
and the hypercharge of the left (right) sfermion, respectively.
We have $Y=2(Q_f - T_3)$ where $Q_f$ is the electric charge. 
For completeness, the sfermion quantum numbers are listed in \tab{quantums}.

\begin{table}
\begin{center}
\begin{tabular}{c|c|c|c|c}
    & $T_{3L}$ & $Q_f$ & $Y_L$ & $Y_R$\\
\hline
  $\st$ & $\phantom{-}1/2$ & $\phantom{-}2/3$ & $1/3$ & $\phantom{-}4/3$ \\
\hline
  $\sb$ & $-1/2$ & $-1/3$ & $1/3$ & $-2/3$ \\
\hline
  ~$\stau$~ & $-1/2$ & $-1$ & $-1$ & $-2$ \\
\end{tabular}
\caption{Isospin, electric charge and hypercharges of stops, sbottoms and staus.}
\label{tab:quantums}
\end{center}
\end{table}

Notice that in the CP-conserving case the pseudoscalars 
only couple to $\tilde f_1^{}\tilde f_2^{}$ combinations and hence 
$g^P_{l11}=g^P_{l22}=0$ in Eqs.~(\ref{eq:gPljkst}) and (\ref{eq:gPljksb}); 
moreover $g^S_{i12}=g^S_{i21}$ and $g^P_{l12}= -g^P_{l21}$. 
Notice also that in the  CP-violating case, the scalar and pseudoscalar 
Higgs states will mix to mass eigenstates $h_{1...5}^0$ similar to 
the MSSM case.
The couplings of the charged Higgs bosons to sfermions are the same as 
in the MSSM.

\subsection{Sfermion--neutralino interaction}

The sfermion interaction with neutralinos has the same form as 
in the MSSM, 
\begin{equation}
  {\cal L}_{f\ti f\nt} = g\,\bar f\,( 
       a^{\,\ti f}_{in} P_R^{} + b^{\,\ti f}_{in} P_L^{} 
  )\,\nt_n\,\ti f_i^{} + {\rm h.c.}\,.
\end{equation}
The only difference is the addition of the singlino state $\ti S$: 
\begin{equation}
   \nt_n = N_{n1}\ti B + N_{n2}\ti W + 
           N_{n3}\ti H_d + N_{n4}\ti H_u + N_{n5}\ti S 
\end{equation}
with $n=1...5$, $N$ the matrix diagonalizing the $5\times 5$ neutralino 
mass matrix in the basis $(\ti B,\ti W,\ti H_d,\ti H_u,\ti S)$: 
\begin{equation}
  {\cal M}_N = 
  \left( \begin{array}{ccccc}
  M_1  &  0  &  -m_Z s_W c_\b  &  m_Z s_W s_\b  &  0 \\
  0  &  M_2  &  m_Z c_W c_\b  &  -m_Z c_W s_\b  &  0 \\
  -m_Z s_W c_\b  &  m_Z c_W c_\b   &  0  &  -\muff  &  -\l v s_\b \\
   m_Z s_W s_\b  &  - m_Zc_W s_\b  &  -\muff  &  0  &  -\l v c_\b \\
  0 & 0 & -\l v s_\b & -\l v c_\b & 2\kappa s  
  \end{array}\right)\,, 
\label{eq:ntmassmat}
\end{equation}
with $s_W=\sin\theta_W$, $c_W=\cos\theta_W$, $s_\b=\sin\b$, $c_\b=\cos\b$, and
\begin{equation}
  N^*{{\cal M}_N} N^\dagger =
  {\rm diag}(\mnt{1},\,\mnt{2},\,\mnt{3},\,\mnt{4},\,\mnt{5})\,.
\end{equation}
We can hence use the couplings $a^{\,\ti f}_{in}$ and $b^{\,\ti f}_{in}$ 
as given in \cite{Bartl:2003pd,Gajdosik:2004ed} with the neutralino index 
running from 1 to 5 instead from 1 to 4. It is worth noting that  
the couplings between sfermions and singlinos only occur via the 
neutralino mixing. This is in contrast to the Higgs couplings where 
additional terms originating from Eq.~\eq{VF} are present in the 
sfermion--singlet interaction. 
Note also that \cite{Ellwanger:2004xm} and \cite{Belanger:2005kh} 
use the basis $(\ti B,\ti W,\ti H_u,\ti H_d,\ti S)$ for the neutralino system.
Therefore, the indices 3 and 4 of the neutralino mixing matrix $N$ 
need to be interchanged when comparing \cite{Ellwanger:2004xm,Belanger:2005kh} 
and this paper.

\section{Numerical results}

We have implemented all 2-body sparticle decays of the NMSSM in the 
{\tt SPheno}~\cite{Porod:2003um} package. 
The Higgs sector of the NMSSM is calculated with 
{\tt NMHDECAY}~\cite{Ellwanger:2004xm} linked to {\tt SPheno}.
For our discussion of sfermion decays in the NMSSM, we choose the 
benchmark scenario~3 of \cite{Belanger:2005kh}
which is characterized by
\begin{eqnarray}
&	\lambda=0.4,\quad
	\kappa=0.028,\quad
	\tan\b=3,\quad
	\muff=\lambda s = 180\gev, & \nonumber\\
&	A_\lambda=580 \gev,\quad
	A_\kappa=-60 \gev,\quad
	M_2 = 660\gev, & 
\label{eq:dm03par}
\end{eqnarray}
with $M_1=(g_1/g_2)^2 M_2\simeq 0.5 M_2$ by GUT relations 
as an illustrative example. 
This leads to a light $\nt_1$ with a mass of 35~GeV which is to 87\% 
a singlino. The $\nt_2$ weights 169~GeV and is dominantly a higgsino. 
Moreover, we have a light scalar Higgs with a mass of 
$m_{H_1}=36$~GeV and a light pseudoscalar with $m_{A_1}=56$~GeV,  
both being almost pure singlet states and thus evading the LEP bounds. 
$H_2^0$, $H_3^0$, and $A_2^0$ are $SU(2)$ doublet fields similar 
to $h^0$, $H^0$ and $A^0$ in the MSSM.
The relic density in this scenario is 
$\Omega h^2=0.1155$~\cite{Belanger:2005kh}.

In this letter, 
we are interested in the decays $\ti f_2^{}\to\ti f_1^{} H_1^0$, 
$\ti f_2^{}\to\ti f_1^{} A_1^0$, and $\ti f_{1,2}^{}\to f\nt_1$, 
with $\ti f=\st,\,\sb,\,\stau$. 
In order to see the relevance of these decays, 
we perform a random scan over the parameters of the third generation, 
$M_{\ti Q_3}$, $M_{\ti U_3}$, $M_{\ti D_3}$, 
$M_{\ti L_3}$, $M_{\ti E_3}$, $A_t$, $A_b$, $A_\tau$. 
The sfermion mass parameters are varied between 100--800~GeV, 
and the trilinear couplings in their whole possible range 
allowed by the absence by charge or colour breaking minima.
We compute the mass spectrum and the branching ratios at each scan point, 
accepting only points which pass the experimental bounds from LEP 
(the bounds from the LEP Higgs searches are fully   
implemented in {\tt NMHDECAY}~\cite{Ellwanger:2004xm}). 
Owing to radiative corrections, the mass of $H_2^0$ 
varies between $\sim 100$~GeV and 117~GeV in the scan. 
The effect on the other quantities, in particular $m_{H_1}$,  
$m_{A_1}$ and $\Omega h^2$, is negligible. 

As a sideremark we note that renormalization-group (RG) arguments 
can be used to set lower bounds on the $SU(2)$ doublet sfermion masses. 
Requiring, for example, that $m_{\tilde f_L}^2$ remain positive all the 
way up to the GUT scale implies $m_{\tilde f_L} \gsim 0.9 M_2$ 
for the first and second generation at the weak scale. 
The corresponding bounds for the third generation are much lower 
because the Yukawa couplings contribute to the RG running with 
opposite sign as the $SU(2)$ gauge coupling. 
We refrain, however, from imposing any such RG-inspired constraint 
in our analysis for two reasons: firstly 
because it is our aim to discuss the weak-scale phenomenology in the most 
general way using just one illustrative benchmark scenario, and secondly 
because the actual scale of SUSY breaking is unknown and may well be much 
lower than $M_{GUT}$ (see e.g.~the NMSSM variants of the models presented 
in \cite{Allanach:2001qe}).

\begin{figure}[t]
\begin{center} \setlength{\unitlength}{1mm}
\begin{picture}(150,140) 
\put(2,2){\mbox{\epsfig{figure=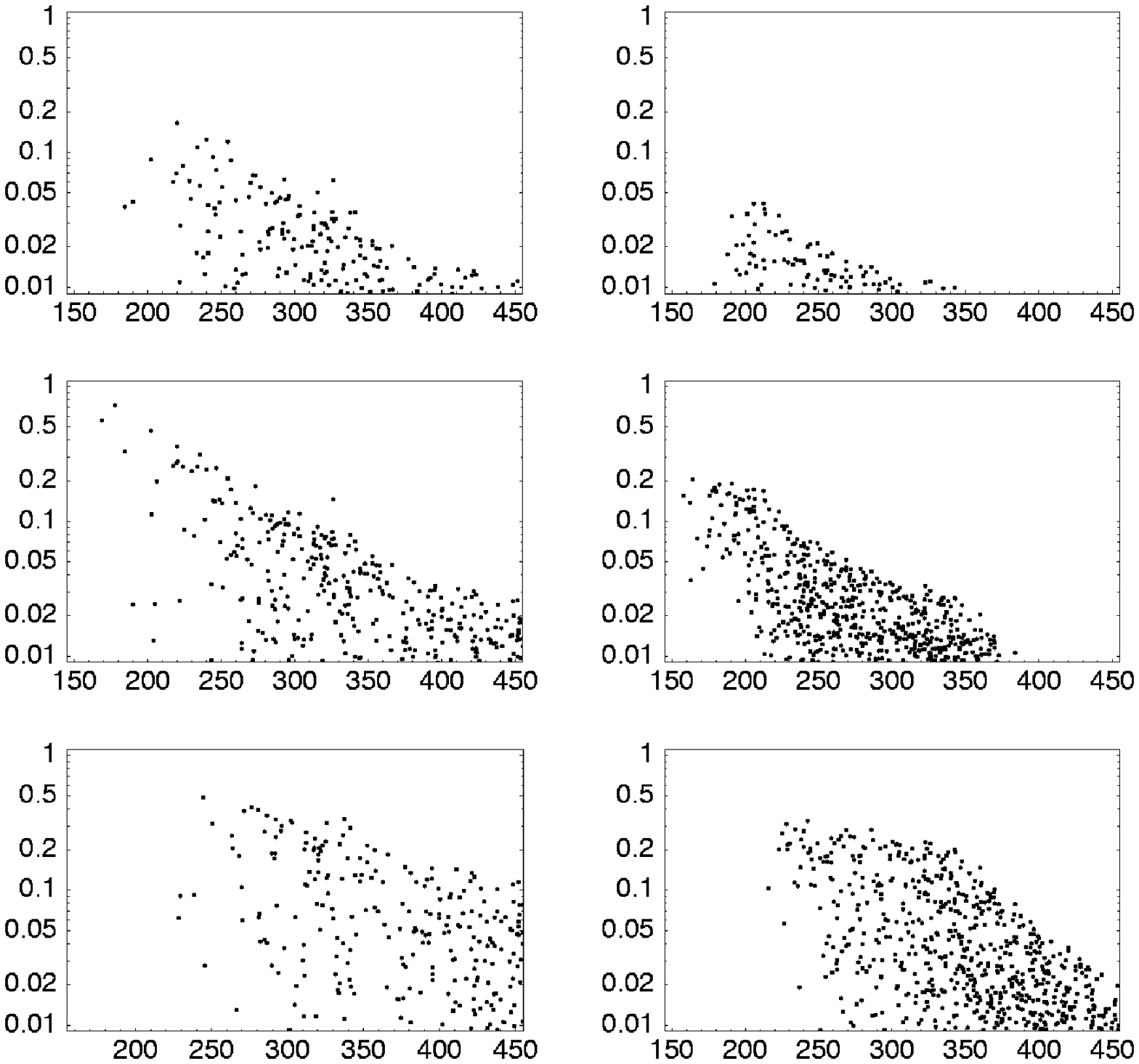,width=15cm}}}
\put(33,0){$\msb{2}$~[GeV]}
\put(33,48){$\msb{2}$~[GeV]}
\put(33,95){$\msb{2}$~[GeV]}
\put(108,0){$\mstau{2}$~[GeV]}
\put(108,48){$\mstau{2}$~[GeV]}
\put(108,95){$\mstau{2}$~[GeV]}
\put(-1,14){\rotatebox{90}{ BR($\sb_2\to\sb_1 H_2^0$) }}
\put(-1,60){\rotatebox{90}{ BR($\sb_2\to\sb_1 H_1^0$) }}
\put(-1,106){\rotatebox{90}{ BR($\sb_2\to\sb_1 A_1^0$) }}
\put(75,14){\rotatebox{90}{ BR($\stau_2\to\stau_1 H_2^0$) }}
\put(75,60){\rotatebox{90}{ BR($\stau_2\to\stau_1 H_1^0$) }}
\put(75,106){\rotatebox{90}{ BR($\stau_2\to\stau_1 A_1^0$) }}
\end{picture} 
\end{center}
\caption{Branching ratios of $\sb_2$ (left) and $\stau_2$ (right) decays 
into Higgs bosons $A_1^0,H_1^0,H_2^0$ for scenario~3 of \cite{Belanger:2005kh}, 
c.f.\ Eq.~\eq{dm03par}, as function of the mass of the decaying particle. 
\label{fig:BR_sbotstau_to_Higgs}}
\end{figure}

Let us now discuss the sfermion branching ratios. 
\Fig{BR_sbotstau_to_Higgs} shows scatter plots of the branching ratios 
of $\sb_2$ and $\stau_2$ decays into $A_1^0$, $H_1^0$ and $H_2^0$ 
as function of the heavy sfermion mass, $\msb{2}$ or $\mstau{2}$. 
As can be seen, decays into the singlet 
Higgs bosons $H_1^0$ or $A_1^0$ can have sizable branching ratios provided 
the sfermions are relatively light, $m_{\sb_2,\stau_2}\lsim 400$ GeV.  
This feature can easily be understood from the partial width
of the decay of a heavier sfermion into a lighter one plus a massless
singlet. For sbottoms we have, for instance, 
\begin{eqnarray}
  \Gamma(\sb_2\to\sb_1 S) = \frac{c}{32\pi} h_b^2\lambda^2
  \left(\frac{v_u}{\msb{2}}\right)^2\,
  \left[ 1 - \left(\frac{\msb{1}}{\msb{2}}\right)^2 \right] \,\msb{2} \,,
\end{eqnarray}
which is suppressed by a factor $(v_u/m_{\sb_{2}})^2$ for heavy sbottoms. 
The factor $c$ is $c=\cos^2 2\theta_{\tilde b}$ for the scalar
and $c=1$ for the pseudoscalar singlet. 
Analogous expressions hold for staus with $b\to \tau$ and  
for stops with $b\to t$ and $v_u\to v_d$.
Decays into the $SU(2)$ doublet Higgs $H_2^0$ can also have large branching 
fractions, provided the splitting of the sfermion mass eigenstates is large 
enough. 
Here note that for the parameter choice Eq.~\eq{dm03par}, $A_1^0$ and $H_1^0$ 
decay predominantly into $b\bar b$ and may hence only be distinguished by 
the different $b\bar b$ invariant masses. The $H_2^0$ on the other hand, 
decays to about 60\% into $H_1^0H_1^0$ and hence into a $4b$ final state.
Both signatures, the one from decay into a $H^0_1$ or $A^0_1$ leading
to $b \bar{b}$ with small invariant mass as well as the 4$b$'s from
the decay into $H^0_2$, are distinct from the usual MSSM case.

\begin{figure}[t]
\begin{center} \setlength{\unitlength}{1mm}
\begin{picture}(150,94) 
\put(2,2){\mbox{\epsfig{figure=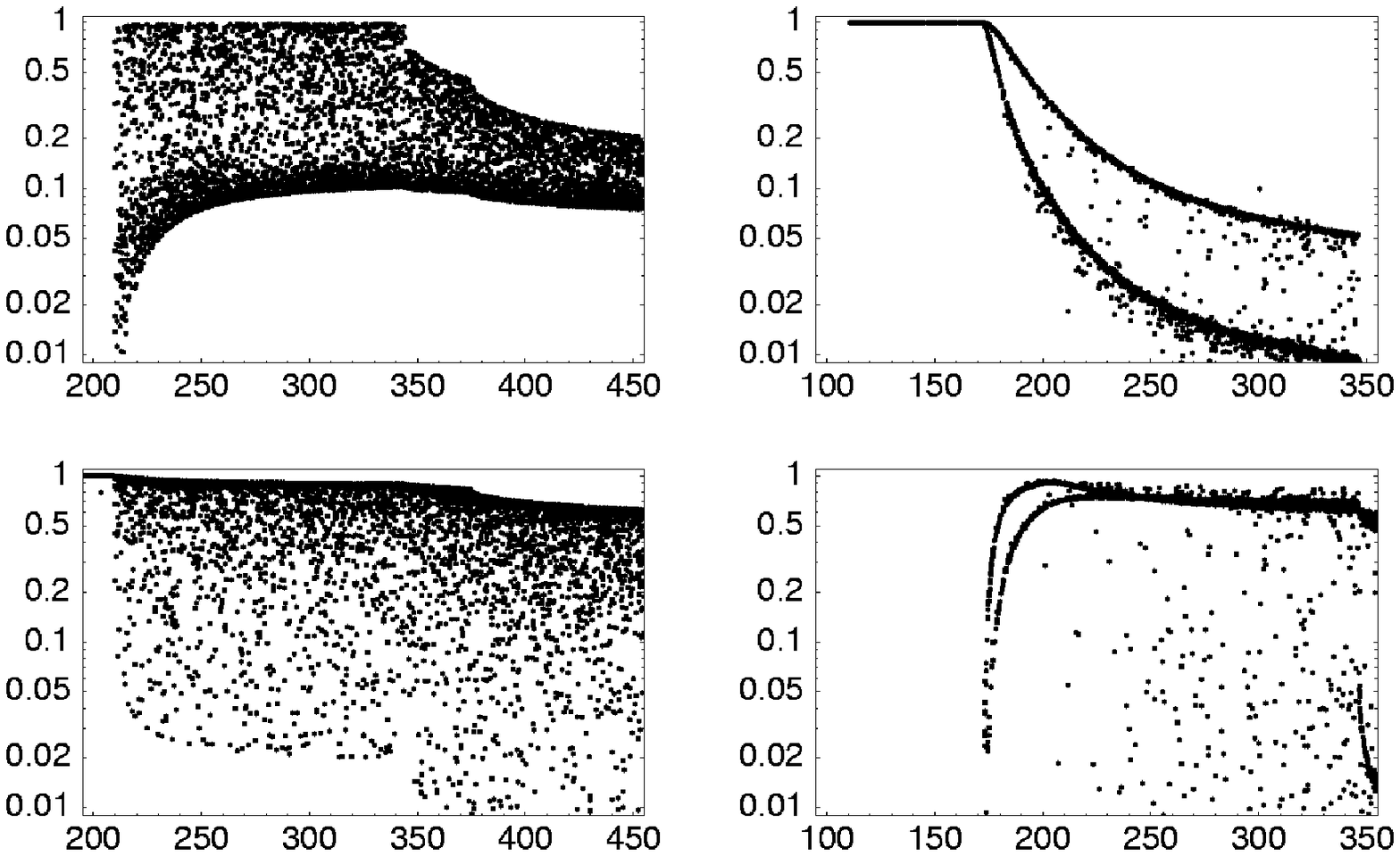,width=15.2cm}}}
\put(34,0){$\mst{1}$~[GeV]}
\put(34,48){$\mst{1}$~[GeV]}
\put(110,0){$\msb{1}$~[GeV]}
\put(110,48){$\msb{1}$~[GeV]}
\put(-1,60){\rotatebox{90}{ BR($\st_1\to t\nt_1$) }}
\put(-1,13){\rotatebox{90}{ BR($\st_1\to b\ch_1$) }}
\put(75,60){\rotatebox{90}{ BR($\sb_1\to b\nt_1$) }}
\put(75,13){\rotatebox{90}{ BR($\sb_1\to b\nt_2$) }}
\end{picture} 
\end{center}
\caption{Branching ratios of $\st_1$ decays into $t\nt_1$ and $b\ch_1$ (left) 
and of $\sb_1$ decays into $b\nt_1$ and $b\nt_q$ (right) for 
the parameters of Eq.~\eq{dm03par}. 
\label{fig:BR_stop_to_chi0}}
\end{figure}

\begin{figure}[t]
\begin{center} \setlength{\unitlength}{1mm}
\begin{picture}(150,46) 
\put(1,2){\mbox{\epsfig{figure=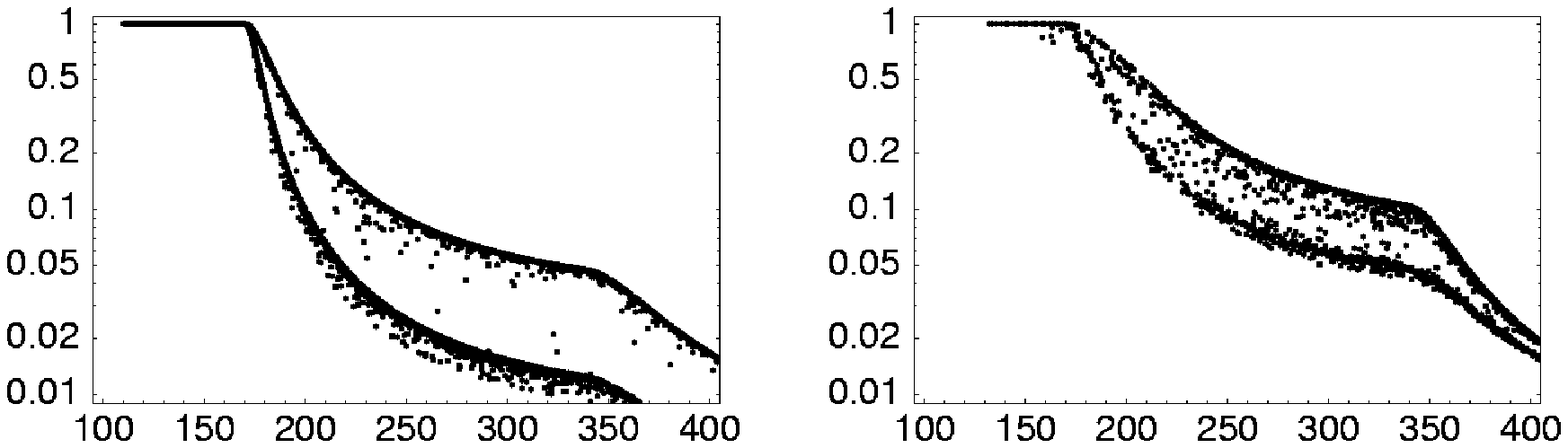,width=15.4cm}}}
\put(34,-1){$\mstau{1}$~[GeV]}
\put(110,-1){$\mstau{2}$~[GeV]}
\put(-1.5,12){\rotatebox{90}{ BR($\stau_1\to \tau\nt_1$) }}
\put(75,12){\rotatebox{90}{ BR($\stau_2\to \tau\nt_1$) }}
\put(44,28){$\stau_1\sim\stau_L$}
\put(21,15){$\stau_1\sim\stau_R$}
\put(122,35){$\stau_2\sim\stau_R$}
\put(105,20){$\stau_2\sim\stau_L$}
\end{picture} 
\end{center}
\caption{Branching ratios of $\stau_1\to \tau\nt_1$ (left) and 
$\stau_2\to \tau\nt_1$ (right) decays for 
the parameters of Eq.~\eq{dm03par}. 
\label{fig:BR_stau_to_chi0}}
\end{figure}

We next turn to sfermion decays into singlinos. 
\Fig{BR_stop_to_chi0} shows scatter plots of the branching ratios 
of $\st_1$ decays into $t\nt_1,\,b\ch_1$, and of $\sb_1$ decays into 
$b\nt_{1,2}$.
As expected, $\st_1\to b\ch_1$ and $\sb_1\to b\nt_2$  
($\ch_1$ and $\nt_2$ being mostly doublet higgsinos) are in general 
the dominant modes if kinematically allowed. Nevertheless, as 
can be seen from the upper two plots of \fig{BR_stop_to_chi0}, decays 
into the singlino LSP can have sizable branching fractions, even if other 
decay modes are open. The size of the branching ratio into the singlino
is mainly governed by its admixture from the doublet higgsinos, 
{\it i.e.} by the size of the $\lambda$ parameter. 
Similar features appear also in the stau decays as illustrated 
in \fig{BR_stau_to_chi0}. 
The pattern of $\stau_1$ is quite similar to that of $\sb_1$, 
with $\gsim10\%$ branching ratio into the singlino for 
$\mstau{1}\lsim 200$--250~GeV, depending on the $L/R$ character of the stau. 
For the $\stau_2$, the branching ratio of the decay into the singlino is 
even more important. In fact it can be 10--50\% over a large part of the 
parameter space even if the decay into $\nt_2$ is open. 

In this context note also the $\ch_1$ and $\nt_2$ will cascade 
further into the singlino LSP, see {\it e.g.}\  
\cite{Franke:1995tf,Ellwanger:1997jj,Choi:2004zx}.
This is quite distinct from the MSSM, where the singlino is absent and 
all decay chains end in what in our case is the $\nt_2$. 
Obviously, this also affects cascade decays of squarks of the 
first and second generation (and likewise of gluinos), since in case of a 
singlino LSP there is one more step in the chain as compared to the MSSM.  
At the LHC, a singlino LSP can in fact lead to similar signatures  
as a gravitino or axino LSP. The presence of light scalar or pseudoscalar 
Higgs bosons in the decay chains may be a way to distinguish the NMSSM from 
other scenarios. If $\lambda$ is large enough the decays $\ti f_i^{}\to f\nt_1$ 
with $\ti f_i^{}\not=\,$NLSP may also be used for discrimination, 
since the corresponding $\ti f_i^{}$ decays into gravitino or axino would 
not occur.
We will discuss this in more detail in a forthcoming paper.

\begin{figure}[t]
\begin{center} \setlength{\unitlength}{1mm}
\begin{picture}(150,46) 
\put(2,-4){\mbox{\epsfig{figure=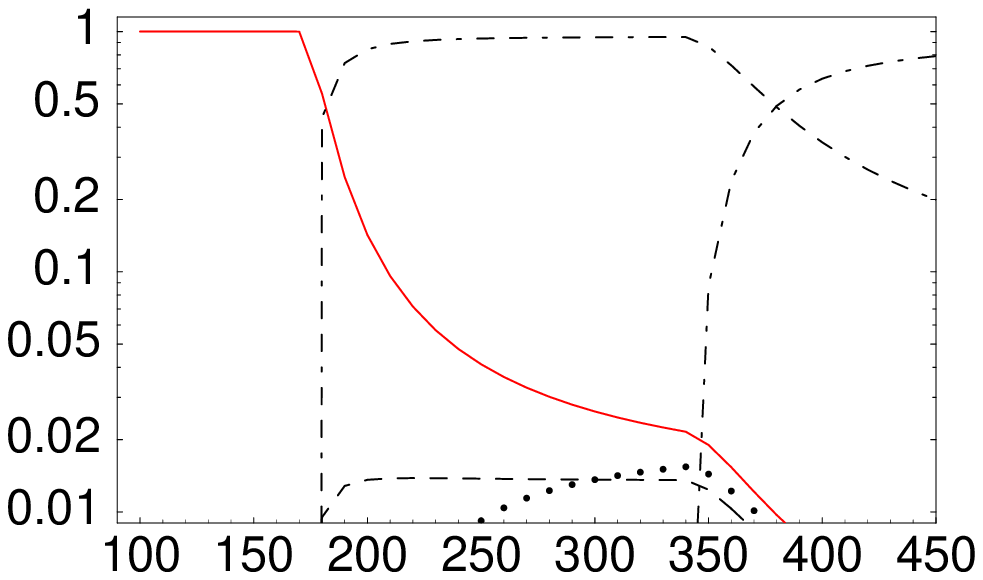,width=7.cm}}}
\put(81,-4){\mbox{\epsfig{figure=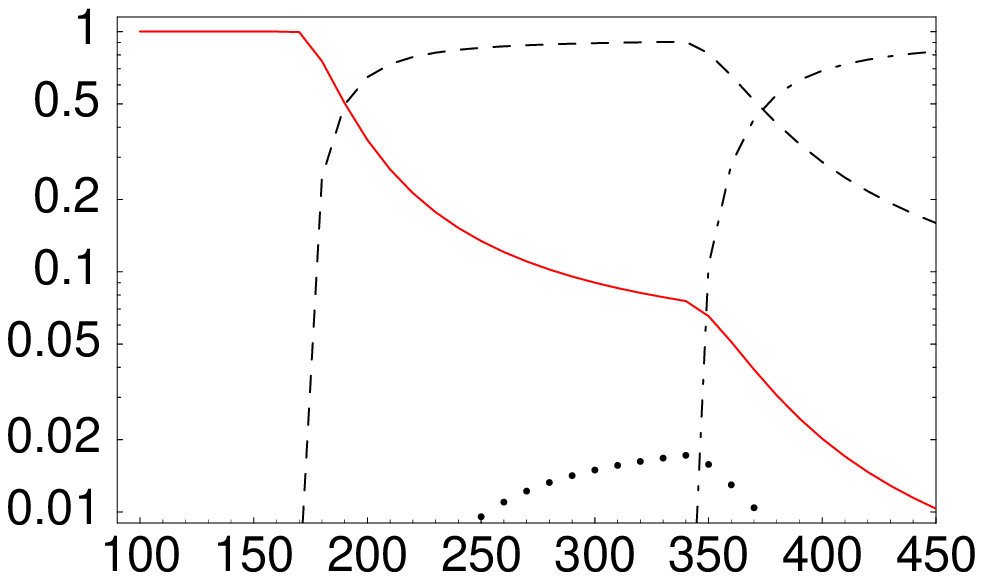,width=7.cm}}}
\put(30,-1.5){$m_{\tilde e_L}$~[GeV]}
\put(109,-1.5){$m_{\tilde e_R}$~[GeV]}
\put(-3.5,18){\rotatebox{90}{ BR\,($\tilde e_L$) }}
\put(75,18){\rotatebox{90}{ BR\,($\tilde e_R$) }}
\end{picture} 
\end{center}
\caption{Branching ratios of $\tilde e_L$ (left) and 
  $\tilde e_R$ (right) decays for the parameters of Eq.~\eq{dm03par}. 
  The full, dashed, dotted, dash-dotted,  and dash-dash-dotted
  lines are for the decays into $e\nt_1$, $e\nt_2$, $e\nt_3$, 
  $e\nt_4$, and $\nu \tilde \chi^-_1$, respectively. The branching ratios 
  for smuons $\tilde\mu_{L,R}$ are the same as for the selectrons.}
\label{fig:BR_selectron_to_chi0}
\end{figure}

Last but not least we consider the decays of sfermions of the  
first two generations.  Owing to the small Yukawa couplings,  
decays into Higgs bosons are negligible in this case. 
Decays into singlinos can, however, be important. 
As an example, \fig{BR_selectron_to_chi0} 
shows the branching ratios of $\ti e_{L,R}^{}$ (being the same as 
those of $\ti\mu_{L,R}^{}$) for the scenario of Eq.~\eq{dm03par}. 
For $m_{\ti e}\simeq 180$--370 GeV, the decays 
$\ti e_L^{}\to\nu_e\ti\x_1^-$ and $\ti e_R^{}\to e\nt_2$ clearly dominate, 
giving a 2-step cascade decay into the singlino LSP. 
Nevertheless even in this case the decays $\ti e_{L,R}^{}\to e\nt_1$ 
have sizable rates, being of the order of 10\% for $\tilde e_R$.
The reason is that the relative importance of the decays into charginos 
and neutralinos is determined by the gaugino components of the these 
particles, and 
$\ch_1$ and $\nt_{2,3}$ are mainly higgsino-like in our scenario. 
Only when the decay into $\nt_4$, which is mainly a bino, gets
kinematically allowed  
the direct decays into the singlino become negligible.
Also for the decays of up and down squarks 
into singlinos we find branching ratios of ${\cal O}$(1--10)\%. 
The implications for collider phenomenology will 
be discussed in detail elsewhere.

\section{Conclusions}

We have discussed the decays of sfermions in the NMSSM.
We have shown that for stops, sbottoms and staus,
in addition to the decay modes already present in the MSSM, 
decays into light singlet Higgs bosons, 
$\ti f_2^{}\to \ti f_1^{}+A_1^0,H_1^0$, 
as well as decays into a singlino LSP,
$\ti f_i^{}\to f\nt_1$ ($i=1,2$) with $\nt_1\simeq\ti S$, 
can be important. 
This is in particular the case for light sfermions.
The presence of these decay modes modifies the signatures of stop, 
sbottom and stau events as compared to the MSSM. 
Also for first and second generation sfermions it turned out that 
the decays into a singlino LSP can be quite important. 
Even if other decay modes are open,  $\ti f_i^{}\to f\nt_1$ with 
$\nt_1\simeq\ti S$ can have ${\cal O}(10\%)$ branching ratio.  
Moreover, decays (of any SUSY particle) into singlinos or singlet Higgs 
bosons may significantly influence cascade decays of squarks and gluinos 
at the LHC. In particular in case of a singlino LSP, there is one more 
possible step in the decay chain than in the MSSM.
A singlino LSP in the NMSSM can in fact lead to similar signatures 
at the LHC as a gravitino or axino LSP in the MSSM. The decays discussed 
in this letter may help discriminating these scenarios.

\section*{Acknowledgments}

We like to thank J.F.~Gunion and C.~Hugonie for useful discussions 
on the NMSSM and on {\tt NMHDECAY}. 
We also thank G.~B\'elanger and A.~Pukhov for comparisons with the 
NMSSM version of {\tt micrOMEGAs}. 
The work of S.K. is financed by an APART grant of the Austrian 
Academy of Sciences. 
W.P.~is supported by a MCyT Ramon y Cajal contract,
by the Spanish grant BFM2002-00345, by the
European Commission Human Potential Program RTN network
HPRN-CT-2000-00148  and 
partly by the Swiss 'Nationalfonds'.


\end{document}